\documentclass[conference]{IEEEtran}

\IEEEoverridecommandlockouts
\usepackage{cite}
\usepackage{amsmath,amssymb,amsfonts}
\usepackage{algorithmic}
\usepackage{graphicx}
\usepackage{textcomp}
\usepackage{xcolor}
\def\BibTeX{{\rm B\kern-.05em{\sc i\kern-.025em b}\kern-.08em
    T\kern-.1667em\lower.7ex\hbox{E}\kern-.125emX}}
    
 \DeclareGraphicsExtensions{.pdf,.png,.jpg}
 \graphicspath{{figs/}}
 
\begin{document}
	
\title{A Digital Forensics Investigation of a Smart Scale IoT Ecosystem}

\author{\IEEEauthorblockN{George Grispos\IEEEauthorrefmark{1}, Frank Tursi\IEEEauthorrefmark{1}, Kim-Kwang Raymond Choo\IEEEauthorrefmark{2}, William Mahoney\IEEEauthorrefmark{1} and William Bradley Glisson\IEEEauthorrefmark{3}} 

\IEEEauthorblockA{\IEEEauthorrefmark{1}School Interdisciplinary Informatics, University of Nebraska--Omaha, USA\\ Email: ggrispos@unomaha.edu, ftursi@unomaha.edu, wmahoney@unomaha.edu } 

\IEEEauthorblockA{\IEEEauthorrefmark{2}Department of Information Systems and Cyber Security, University of Texas at San Antonio, USA\\	Email: raymond.choo@fulbrightmail.org}  

\IEEEauthorblockA{\IEEEauthorrefmark{3}Cyber Forensics Intelligence Center, Sam Houston State University, USA\\ Email: glisson@shsu.edu}}

\maketitle

\begin{abstract}
The introduction of Internet of Things (IoT) ecosystems into personal homes and businesses prompts the idea that such ecosystems contain residual data, which can be used as digital evidence in court proceedings. However, the forensic examination of IoT ecosystems introduces a number of investigative problems for the digital forensics community. One of these problems is the limited availability of practical processes and techniques to guide the preservation and analysis of residual data from these ecosystems. Focusing on a detailed case study of the iHealth Smart Scale ecosystem, we present an empirical demonstration of practical techniques to recover residual data from different evidence sources within a smart scale ecosystem. We also document the artifacts that can be recovered from a smart scale ecosystem, which could inform a digital (forensic) investigation. The findings in this research provides a foundation for future studies regarding the development of processes and techniques suitable for extracting and examining residual data from IoT ecosystems. 
\end{abstract}

\begin{IEEEkeywords}

Internet of Things, IoT Forensics, Digital Forensics, Incident Response, Embedded Systems\footnote{Please cite this paper as:  \textit{Grispos, G., F. Tursi, K.K.R Choo, W. Mahoney, and W.B. Glisson (2021) ``A Digital Forensics Investigation of a Smart Scale IoT Ecosystem''. \textit{The 20th IEEE International Conference on Trust, Security and Privacy in Computing and Communications (IEEE TrustCom 2021)}}.}

\end{IEEEkeywords}

\section{Introduction}
\label{intro}

Internet of Things (IoT) devices are increasingly being integrated into everyday life. For example, Statista claim that 13.8 billion IoT devices are currently deployed worldwide, and predict that by 2025 this figure will surpass 30 billion~\cite{Statista21IoT}. IoT devices have found their way into a variety of organizations, hospitals and even private homes, allowing users to share information through cloud computing solutions. As a result, deployed IoT devices are generating, collecting, and transmitting large amounts of information~\cite{DBLP:journals/iotj/LiCSBC19,grispos2021investigating}. 

While the collection and dissemination of this information can help providers deliver better services, there are also potential security and privacy risks. For example, such data could be vulnerable to attacks by cybercriminals and other malicious users~\cite{Flynn20Knock}. These risks are starting to materialize through the number of security incidents impacting IoT devices and the organizations that deploy such devices on their networks. The 2020 Pulse Secure Endpoint and IoT Zero Trust Security Report \cite{Pulse20Report} indicates that 72\% of organizations experienced an IoT security incident in the last year. Further complicating matters, a separate report from Irdeto \cite{Irdeto19Report} suggests that 42\% of IoT attacks results in compromised user or device data. Hence, cybercriminals who maliciously collect information from these devices could cross--link this information from other sources and develop detailed intelligence about individuals and businesses \cite{Irdeto19Report,grispos2019bleeding}.

The increasing use of IoT devices reinforces the importance of digital evidence in various legal situations~\cite{heckmann36medical,maras2020state}. Data produced and stored by IoT devices could provide important evidence in criminal and civil court proceedings. For example, in the United States a suspect was arrested and charged for sexual assault using evidence captured by an IoT camera~\cite{Muri20Wash}. However, while the inclusion of IoT evidence into a variety of legal situations is increasing, researchers have argued that the forensic investigation of IoT devices is unlikely to be straightforward \cite{DBLP:journals/iotj/LiCSBC19,DBLP:journals/access/QuickC18}. One of the biggest concerns raised is the suitability and effectiveness of traditional forensic evidence acquisition approaches in IoT environments~\cite{hegarty2014digital,zawoad2015faiot}. Furthermore, researchers have also highlighted the need for further research to develop and improve forensic tools and processes to assist investigators with device disassembly, data acquisition, and preservation of evidence in IoT devices and their ecosystems \cite{atlam2020internet,wu2019iot}.

An emerging trend within the IoT community is the development of smartphone applications, which act as an interface to their respective IoT device(s) \cite{Flynn20Knock,chao2019enhanced}. The purpose of these applications is to enable the end--user to receive readings or data from an IoT device, and then either store or exchange this information with remote cloud services or other devices. Effectively, these smartphone applications and cloud services, together with the IoT devices, have resulted in the creation of \textit{IoT ecosystems}. The large amount of data generated by IoT devices, in conjunction with increased usage of such data in court cases, supports the idea that the recovery and analysis of evidence from IoT ecosystems will increase in importance in legal settings. Moreover, the increased use and popularity of IoT ecosystems could result in them becoming targets of interest to cybercriminals, including Advanced Persistent Threat (APT) actors. 

In this paper, we present a detailed case study of the iHealth Smart Scale ecosystem. The contributions of this research are three-fold. First, the research provides an empirical demonstration of practical techniques for recovering residual data from different evidence sources within an IoT ecosystem. Second, the research documents the forensic artifacts that can be recovered from a smart scale IoT ecosystem, which could be  useful in a digital forensics investigation. Third, the work provides the foundation for future studies regarding the development of processes and techniques suitable for extracting and examining evidence from IoT ecosystems. 

The remainder of the paper is structured as follows. Section \ref{related} presents the extant literature. Section \ref{method} presents the research methodology, followed by the findings and discussion in Section \ref{results}. Section \ref{conclusions} concludes the paper and presents future research.

\section{Related Literature}
\label{related}

Growing security concerns with IoT devices has resulted in increased research that attempts to enhance a digital forensic investigator's knowledge and ability to conduct investigations of these devices. While the objective of many forensic investigations is to answer the 5 W's and 1 H (what, why, who, when, where, and how), researchers have raised concerns regarding the ability to answer these questions during an investigation of IoT environments, which are expected to be intrinsically harder to investigate as compared to conventional digital devices \cite{DBLP:journals/iotj/LiCSBC19,DBLP:journals/access/QuickC18}. Zawoad and Hasan define IoT forensics where ``the traditional forensic process of identification, collection, organization, and presentation are applied to criminal incidents involving IoT infrastructures" \cite{zawoad2015faiot}. These researchers go on to describe IoT forensics as an amalgamation of investigating end--devices, network forensics, and cloud forensics. Hegarty, et al. \cite{hegarty2014digital} add that IoT ecosystems could provide more digital evidence, as compared to conventional computer systems. Hence, several researchers have attempted to address the challenges associated with investigating IoT environments. 

Wu, et al. \cite{wu2019iot} surveyed individuals to identify and establish current and future research challenges associated with investigating IoT ecosystems. The survey results identified that while IoT forensics is considered a sub--domain of digital forensics, the respondents were undecided about the specific domains that make up IoT forensics. Further, the survey results indicate that urgent research is needed to develop tools and techniques to assist investigators with the preservation and acquisition of data from the cloud, as well as processes to assist investigators in the disassembly of IoT devices \cite{wu2019iot}. Hegarty, et al. \cite{hegarty2014digital} also argue that certain techniques and approaches for investigating traditional desktops and servers become invalidated in IoT environments. Hegarty et al. go to identify several challenges for forensic investigators, including preserving evidence from IoT devices and collecting evidence from aggregated IoT evidence sources. Macdermott, et al. \cite{macdermott2018iot} discuss the forensic challenges posed by IoT ecosystems and argue the applicability of the Association of Chief Police Officers guidelines in such environments. The authors also highlight the challenge due to the fact that potential IoT evidence can come in different formats and is stored in a variety of locations. 

An IoT forensic investigation will likely require the collection of evidence from IoT devices. However, this is also expected to introduce a number of challenges for investigators. Watson and Dehghantanha \cite{watson2016digital} argue that IoT devices are embedded systems that include printed circuit boards, microcontrollers, flash memory, and network components. As a result, Watson and Dehghantanha question whether traditional digital forensics tools and techniques can be used to obtain this data. Zawoad and Hasan counter this argument by indicating that the investigation of IoT devices will result in ``new opportunities for digital forensics investigators'' \cite{zawoad2015faiot}. Therefore, Zawoad and Hasan propose an investigative model called Forensics Aware Internet of Things (FAIoT) to support and guide forensic investigations within IoT ecosystems. While FAIoT includes a centralized evidence repository to assist with the collection and analysis of IoT evidence, the approach was not empirically validated, nor is it clear how extracting evidence from the repository maintains the chain of custody. Oriwoh, et al. \cite{oriwoh2013internet} suggest that IoT devices are likely to be networked within `zones' such as internal networks, remote servers, and the cloud. As a result, they proposed the idea of investigating these three zones collectively in an attempt to maximize the amount of digital evidence recovered from IoT devices. While the authors also propose a hypothetical scenario, the `zone approach' is not empirically validated to determine the actual recoverable evidence from each of the zones \cite{oriwoh2013internet}. 

From a medical device perspective, researchers \cite{grispos2017medical,grispos2020cyber} have argued that  the forensic investigation of medical devices is unlikely to be straightforward and that further research is needed to identify residual data from these devices, which could be of use to forensic investigators. For example, Ellouze, et al.~\cite{ellouze2016forensic} investigate implantable medical devices, with the aim of identifying the digital evidence sources available during an investigation of these devices. The authors concluded that forensic investigators need to investigate a variety of sources including sensors, wireless nodes and applications to identify what has happened and potentially who is responsible for a crime or incident involving implantable devices. While previous research has identified the challenges associated with conducting forensic investigations of IoT environments, minimal research has empirically examined the digital evidence that can actually be retrieved from different components within an IoT ecosystem, such as a smart scale ecosystem. 

\section{Exploratory Case Study}
\label{method}

At a high--level, this research can be best described as an exploratory case study \cite{oates2005researching}. The hypothesis that guides the case study is that \textit{IoT ecosystems can provide forensically--relevant residual data}. This hypothesis provokes the following research questions: Is it possible to recover metadata about an individual and their use of an IoT device from an IoT ecosystem? If user and device metadata is recoverable from an IoT ecosystem, what does this metadata tell us about a user's activities and device usage? Does the manipulation of components (e.g., smartphone applications and cloud services) within an IoT ecosystem influence the collection of metadata from the ecosystem?

A two--phase approach was used in order to investigate the hypothesis and research questions that extends ideas presented by Miller et al. \cite{miller2019investigating}. The first phase consists of a \textit{black box analysis}, whose purpose is to identify hardware capabilities, data storage locations in the smart scale IoT ecosystem, and potential data access methods. The results of the black box analysis influenced the approach used for the second phase, which involves the implementation of a \textit{grey box strategy}. This strategy was chosen due to the availability of open--source documentation (such as \cite{SGS2020Internal}), describing the smart scale. 
 
The device included in this case study was the iHealth Wireless Body Analysis Scale (model HS6). The scale allows a user to record and store scale readings to either an Android or iOS smartphone application called iHealth MyVitals. Both these applications were included in the case study. This medical--grade scale can be used to track and measure several parameters including weight, visceral fat rating, body water, body mass index, lean mass, bone mass, and daily calorie intake. For the purpose of this experiment, the HS6 scale is considered representative of a class of IoT devices that is likely to contain residual data \cite{Flynn20Knock, DBLP:journals/cem/AlladiCSC20}. The justification for this classification is the ability to undertake readings without a smartphone application, and the presence of a smartphone application feature that allows the user to `pull' readings from the scale. The decision to use the specific scale ecosystem was based on its availability.

\subsection{Black Box Examination}
\label{blackbox}

In order to identify the scale's hardware features, potential data storage locations, and approaches for accessing these storage locations, the following steps were undertaken:

\begin{enumerate}
	\item Identify and retrieve open--source documentation to aid in the disassembly of the scale, in order to access the device's internal components, such as the motherboard.
	
	\item  Conduct a visual examination of the motherboard to identify any Integrated Circuit (IC) chips, in order to determine if these chips store the device's firmware or user data. Visual markings on the chips were used to identify the chip using online sources.
	
	\item The datasheet for each chip was then downloaded from the chip vendor's website. The datasheets provide information such as chip features, storage capacity, and serial connection options. This information provides the ability to locate memory chips on the device's motherboard.  
	
	\item The identification of a memory chip then resulted in the visual confirmation of the chip's form factor, as well as the chip's specific pin configuration.
	
	\item Using the above information, the identified memory chip was then connected to a breakout board containing a FTDI -- FT2232H USB 2.0 IO chip (Fig. \ref{fig:fig1}). The breakout board was then connected via USB to a workstation running Ubuntu Linux (Fig. \ref{fig:fig2}).
	
	\begin{figure}[h!]
		\centering
		\includegraphics[trim=0 0 0 0, width=0.45\linewidth]{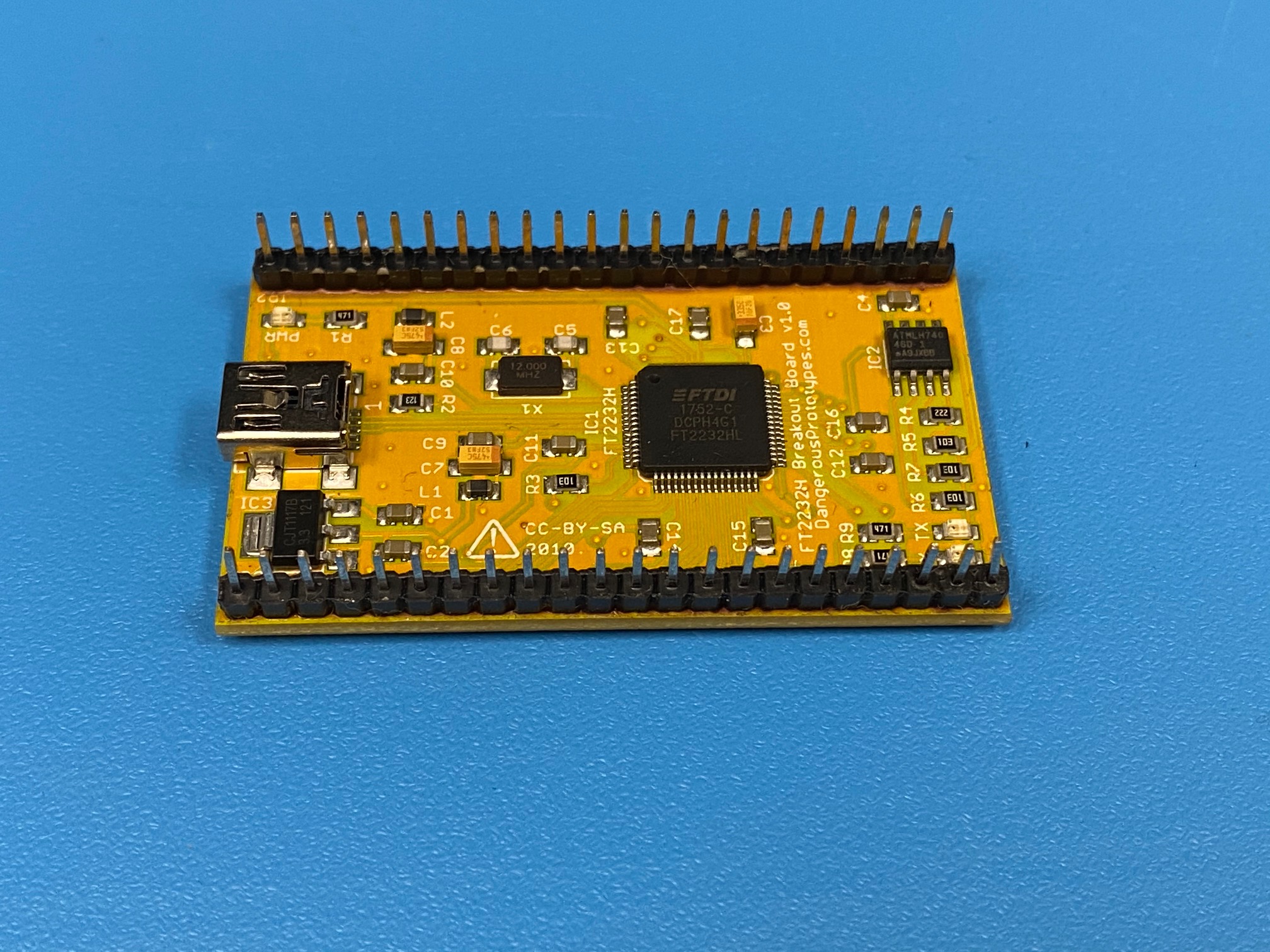}
		\caption{FT2232H Breakout Board}
		\label{fig:fig1}
	\end{figure}
	
	\item Flashrom (version 1.1) was downloaded and installed onto the Ubuntu Linux workstation. After the installation was completed, the following command was executed: \texttt{flashrom -p ft2232 spi:type=2232H, port=A -r  <outputfilename>}. This command reads the memory chip and creates a binary dump of its contents into a local file called \texttt{<outputfilename>}. This process took approximately five minutes to complete. After the extraction was completed, Flashrom successfully verified the contents of the binary dump against the contents of the flash memory chip using a hash algorithm calculation. 
	
	\item To confirm that the scale is indeed recording data, the scale was reassembled and used for a single reading. After taking the reading, the scale was again disassembled and the process described in steps five and six were repeated in order to acquire a second binary dump.The contents of the second binary dump are then compared to the contents of the first binary dump using the Linux commands \texttt{diff} and \texttt{xxd}. In this way, the difference between the two dumps would be the single reading, which is then saved to a separate binary file for analysis.
	
	\begin{figure}[ht]
		\centering
		\includegraphics[trim=0 0 0 0, width=0.45\linewidth]{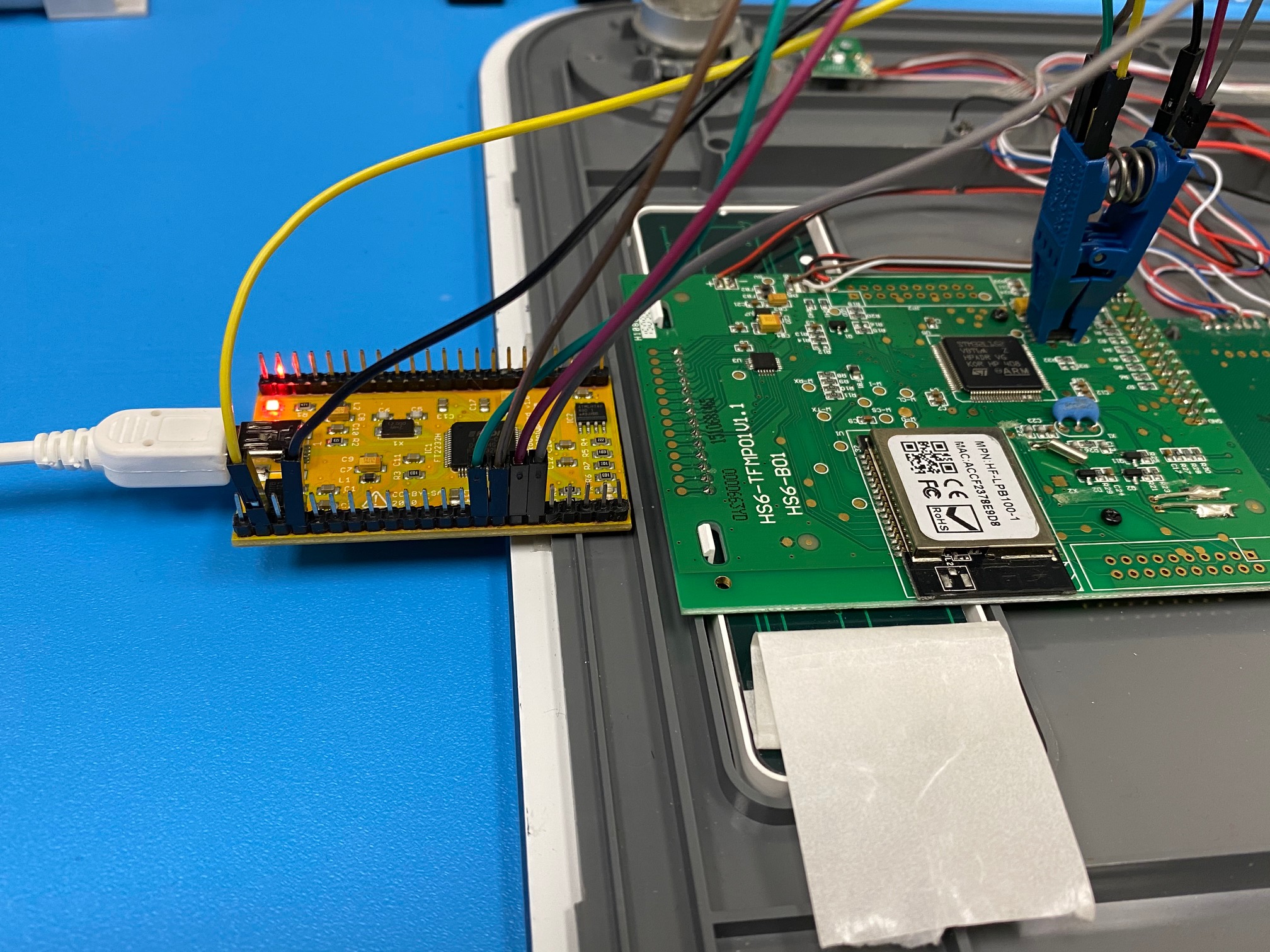}
		\caption{Data Acquisition from HS6 Scale}
		\label{fig:fig2}
	\end{figure}

\end{enumerate}

The results of the black box examination are used in the implementation of the grey box strategy.

\subsection{Grey Box Examination}

In order to investigate evidence generation within the scale ecosystem, one pretest and four treatment with associated posttest experiments were undertaken \cite{oates2005researching}. In addition to the HS6 scale, the ecosystem evaluated in this experiment also included two smartphones running the iHealth MyVitals mobile application on both Android and iOS operating systems, along with a desktop computer used to access the iHealth cloud portal through a web browser. The Android smartphone application used in this experiment was MyVitals version 3.8.1, while the iOS smartphone application was MyVitals version 3.10.1. The grey box examination consists of the following seven--steps:

\begin{enumerate}
	\item The smartphones were reset in order to return the devices to their respective `factory settings' and a desktop computer was used to create an email account for the purpose of the experiment.  The smartphones were then powered--on and a test email account was used to complete the initial device setup. The smartphones were connected to the Internet and the respective `App' store was used to download and install the MyVitals application. The MyVitals application was executed and a new user profile was created. It should be noted that the same test user profile was created on both the Android and iOS smartphones, since a user could use either or both smartphones when using the scale. 
		
	\item After the test user profile were created, each smartphone was `paired' with the HS6 scale, which prompted the recording of an initial scale measurement. This involved observing the measurement as displayed on the scale and then refreshing each smartphone application. The initial measurement was visible on both the Android and iOS application interfaces, which confirms that the scale and smartphones are successfully `paired'. The test email address was then used to access the iHealth cloud portal in order to confirm that the initial measurement was also under the test user's profile. At this point, the setup of the scale ecosystem was considered complete.

	\item The scale was then used once a day, for three days. This consisted of stepping on to the scale, reading the measurement displayed on the scale and on each smartphone applications' interface, and then recording this information. The date and time were also noted for each measurement. According to the iHealth website~\cite{Ihealth2016Fat}, a user is required to stand barefoot on the scale in order for body fat to be measured. In order to determine which measurements record the body fat, the first two daily measurements were with shoes, and the third day measurement was without shoes. After each scale measurement was undertaken, both smartphone applications and the cloud portal account were refreshed to visually confirm the particular measurement was displayed on the respective interface. Table \ref{tab:TestData} presents the measurements displayed on the scale interface during the experiment.
	
	\begin{table}
		\begin{center}
			\begin{tabular}{| l | l | l | l | l | }
				\hline			
				\multicolumn{1}{|c|}{\textbf{Measure}} & \multicolumn{1}{|c|}{\textbf{Day 1}} & \multicolumn{1}{|c|}{\textbf{Day 2}} & \multicolumn{1}{|c|}{\textbf{Day 3}} & \multicolumn{1}{|c|}{\textbf{Day 4}} \\ \hline	
				
				Date & 5/15/2019 & 5/16/2019 & 5/17/2019 & 05/18/2019 \\ \hline
				Time & 4:27 PM & 4:55 PM & 6:28 PM & 3:46 PM \\ \hline
				Weight & 89.6 kg & 88.9 kg & 90.4 kg & 90.2 kg \\ \hline
				BMI  & 25.9 & 25.7 & 26.1 & 26.1  \\ \hline
				Body Fat & 24.4\% & 24.4\% & N/A & 22.8\%  \\ \hline
				
			\end{tabular}
		\end{center}
		\caption{Test Data for Days 1 -- 4}
		\label{tab:TestData}
	\end{table}
	
	\item After the three days, the smartphones were processed using a MSB XRY forensic toolkit (version 3.2) to create forensic images of the smartphones and the HS6 scale was processed using the approach described in steps five and six in Section \ref{blackbox}. The cloud portal was visually examined using the test email address and the password set at the profile creation in order to quantify the measurements stored in the cloud. The results of the cloud analysis, along with the smartphone and scale extractions are grouped as Result Set 1.
	
	\item After the above step, the caches were cleared from the MyVitals applications on both smartphones, and the applications were then deleted from the smartphones. This was done to mimic a real--world anti--forensic scenario. The smartphones and the scale were then processed, and the cloud portal was visually examined. These results are grouped as Result Set 2.
	
	\item The MyVitals applications were reinstalled on both smartphones, and the test credentials were used to recover the test profiles. The smartphone application's interfaces were refreshed and the measurements were visible on the smartphone interfaces. The user's profile was then deleted from the scale using instructions on the iHealth website \cite{removeuser}. The smartphones and the scale were processed, and the cloud portal was visually examined. These results are grouped as Result Set 3.
	
	\item The scale was then used to record a new measurement (Day 4 in Table \ref{tab:TestData}). This measurement was undertaken without the researcher wearing any shoes. The smartphone application interfaces were refreshed, and the Day 4 measurement was visible on both application interfaces. No further actions were undertaken taken between removing the user and the recording of the new measurement. The smartphones and the scale were processed, and the cloud portal was visually examined. These extractions are grouped as Result Set 4.
	
	\item The Android MyVitals application was then used to delete the measurement from Day 4. Both smartphone applications were refreshed to ensure that the reading was no longer visible on the interfaces. The smartphones and the scale were then processed and the cloud portal was visually examined to produce Result Set 5. 
	
\end{enumerate}

The five result sets were then analyzed in order to determine: a) if evidence is stored in each component of the scale ecosystem, b) what evidence can be recovered from each component, and c) the impact of the manipulation on the digital evidence stored within each component. The result sets from Android and iOS smartphones were analyzed using XAMN mobile forensic software, while the scale memory dump was analyzed using a hex editor, and the cloud portal was visually examined.

\subsection{Limitations}
\label{limits}

This research is limited in the following ways. The experiment involved a single instance of a IoT device, an iHealth HS6 scale, acquired from the manufacturer's United States (US) website. It should be noted that there were no visible or documented means to restore the scale to its factory settings. The experiment was undertaken using smartphones that contain mobile carrier software for providers in the US. Finally, the mobile forensic toolkit and smartphones used in this experiment were selected based on availability to the authors. 

\section{Findings and Discussion}
\label{results}

An analysis of Result Set 1, suggests that artifacts concerning the test user can be recovered from the evidence sources within the HS6 scale ecosystem. The following subsections describe the artifacts recovered from the sources. 

\subsection{HS6 Scale Analysis}

\begin{figure}[h]
	\centering
	\includegraphics[trim=0 45 0 0, width=0.9\linewidth]{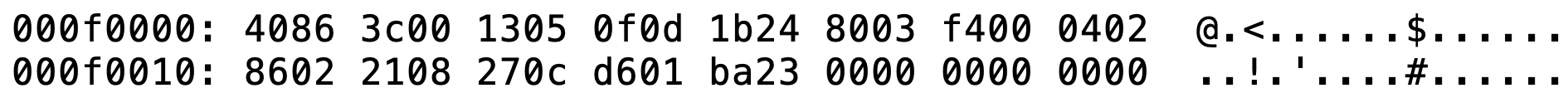}
	\caption{Scale Reading and Decoding from Day 1}
	\label{fig:day1}
\end{figure}

The analysis of the scale's binary dump revealed artifacts related to the test user and their use of the scale. Currently, this information needs to be manually decoded from little--endian hexadecimal to ASCII characters. Unless noted, this translation is needed for all the bytes of information recovered from the scale. Figure \ref{fig:day1} presents the hexadecimal information retrieved from the scale, which represents scale use from Day 1. The following describes how the hexadecimal information can be decoded into ASCII characters.  

The first four bytes (\texttt{40863c00}) correspond to the user identifier, as allocated by the iHealth service, which translate to the decimal value `3966528'. The next four bytes (\texttt{13050f}) correspond to the date when the user undertook a scale reading, in the format (\texttt{dd-mm-yy}), which translates as `15-9-19'. The next three bytes (\texttt{0d1b24}) correspond to the time when a scale reading was undertaken, in the format (\texttt{hh-mm-ss}). To decode this information, the bytes must first be read as big--endian, and then converted to decimal. The time related to the reading is three hours before the local time when the reading was actually undertaken so the decoded time is `13:27:36'. The next two bytes (\texttt{8003}) store the user's weight in floating--point kilograms, which is translated and decoded into decimal as `896', which is `89.6 kg'. The next two bytes (\texttt{f400}) correspond to the user's body fat in floating--point percentage and can be decoded as 244', which is `24.4\%'. The next two bytes (\texttt{0402}) represent the user's body water as a floating-point percentage. Therefore, this value corresponds to `516', which is `51.6\%'. The next two bytes (\texttt{8602}) represent the user's muscle weight stored in floating--point kilograms, which when converted to decimal, is `646', which is `64.6kg'. The next byte (\texttt{21}) represents the user's bone mass stored in floating--point kilograms. In decimal, this value is `33', which is `3.3 kg'. The next byte (\texttt{08}) corresponds to the user's visceral fat rating, which in this scenario, the decimal value is `8'. The next three bytes (\texttt{270cd6}) are unknown, and further investigation is needed. The final three bytes (\texttt{01}), (\texttt{ba}) and (\texttt{23}), represent the user's gender, height in centimeters and age, respectively. Concerning the gender, (\texttt{01}) denotes a `male' user. The user's height is translated as `186 cm', while the age is translated as `35' years old, respectively.
 
\begin{figure}[h]
	\centering
	\includegraphics[trim=0 30 0 0, width=0.9\linewidth]{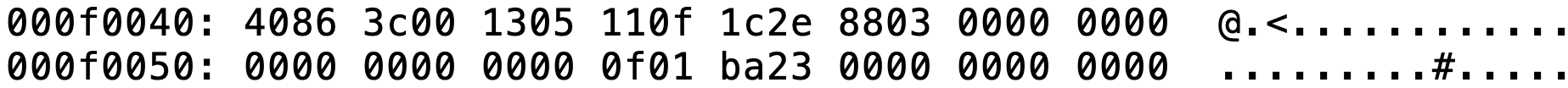}
	\caption{Memory Dump Reading from Day 3}
	\label{fig:day3}
\end{figure}

While the above decoding was successfully applied to the Day 1 and Day 2 memory extractions, the decoding of the Day 3 memory extraction revealed slightly different results, as shown in Figure \ref{fig:day3}. When the scale reading was taken with shoes, the scale did not record and present the body fat percentage. Instead, The hexadecimal value  \texttt{0000} is stored after the user's weight (\texttt{8803}). The hexadecimal value  \texttt{0000} is also used by the scale to store information in the memory location for the user's body water measurement, muscle weight, bone mass, and visceral fat rating.

\subsection{Smartphone Applications Analysis}

The analysis of the Android and iOS smartphone applications revealed that both applications create various artifacts associated with the user and their use of the scale.

\subsubsection{Android Application}

\begin{figure}[h]
	\centering
	\includegraphics[trim=0 0 0 0, width=0.85\linewidth]{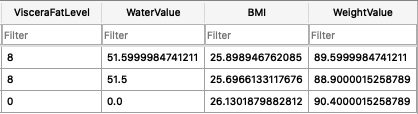}
	\caption{Extract from TB\_WeightOnlineResult Table}
	\label{fig:andapp}
\end{figure}

The Android MyVitals application creates a parent folder called \texttt{iHealthMyVitals.V2}. A variety of artifacts can be recovered from this folder. A subfolder called \texttt{databases} contains a database called \texttt{androidNin.db}, which contains several tables of interest. User--specific metadata can be retrieved from a table called \texttt{TB\_UserInfo}. Information that can be retrieved includes the username, a password hash, the country and timezone they are located, the user's name, their birthday, gender, height, weight, and the user's identification number as allocated by the iHealth service. Information regarding the scale including the device name and identifier, firmware version and the name of the iHealth account(s) that use the particular device can be found in a table called \texttt{TB\_Device}. The third table of interest is called \texttt{TB\_WeightOnlineResult}, which contains the user recordings from the scale. Metadata recovered from this table includes the user's weight, body mass index value, body fat percentage, percentage of body water, visceral fat rating, muscle mass, daily calorie intake, and bone mass. Timestamp information for each reading was also recovered from this table. Figure \ref{fig:andapp} presents extract of the \texttt{TB\_WeightOnlineResult} table showing user data from the first three days of the experiment. In addition to the database, user and device data can also be retrieved from Extensible Markup Language (XML) files. These files can be used to recover the user's email address, the device name, the device's MAC network address, as well as the access token and password hash for the user's account. 

\subsubsection{iOS Application}

The iOS MyVitals application creates a parent folder called \texttt{com.ihealthlabs.iHealth}. Within this folder is a subfolder called \texttt{Documents}, which contains a SQLite database called \texttt{ihealth.sqlite}. This database consists of sixty-four tables; however, only four tables contain metadata related to the experiment.  The following evidence can be recovered from each of these four tables: 

\begin{enumerate}
	\item \texttt{ZUSER} Table -- user metadata including age, birthday, height, weight, email address, username, and location.
	
	\item \texttt{ZSCALETEMPRHINFO} Table -- metadata related to the temperature and humidity at the time of the recording, as well as timestamp information. 
	
	\item \texttt{ZSCALEMEASUREMENT} Table -- scale usage metadata including the user's weight, body mass index, percentage of body fat, body water, muscle mass, daily calorie intake, bone mass, along with timestamp information.
	
	\item \texttt{ZACCESSORYCONNECTLOG} Table -- device metadata including the device name, type, firmware and hardware versions, model number, and device serial numbers.
	
\end{enumerate}

\begin{figure}[ht]
	\centering
	\includegraphics[trim=0 0 0 0, width=0.85\linewidth]{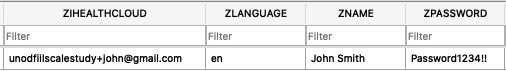}
	\caption{Recovery of Plaintext Password}
	\label{fig:applepass}
\end{figure}

An analysis of the \texttt{ZUSER} table also revealed the recovery of the user's password in plaintext, as shown in Figure \ref{fig:applepass}. 

\subsection{Cloud Portal Analysis}

To access the iHealth cloud portal, a forensic investigator requires the username and password from the user whose activities are under investigation. There are various options for an investigator to obtain this information. First, the investigator could recover the username and password from an artifact stored on another device used to access the iHealth services, as seen in Figure \ref{fig:applepass}. Second, best practices \cite{volonino2006computer} suggest that an investigator can ask the individual for their authentication credentials. This scenario was recently validated in a recent investigation regarding access to Fitbit evidence \cite{snyder15Police}. Third, in certain jurisdictions, a forensic investigator can obtain a legal warrant compelling the individual to provide the required information \cite{volonino2006computer}. Finally, forensic investigators can perform a brute--force cracking attack using the individual's email address and, potentially, a dictionary of potential passwords obtained through analysis of other evidence.

\begin{figure}[!h]
	\centering
	\includegraphics[trim=0 30 0 30, width=0.80\linewidth]{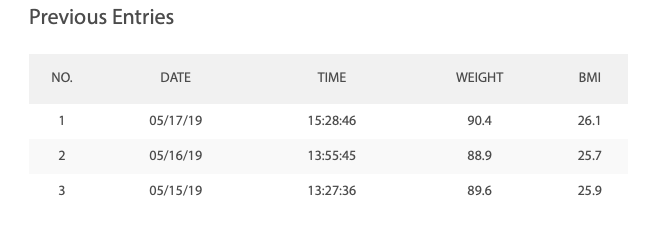}
	\caption{Snapshot of iHealth Cloud Portal}
	\label{fig:portalprevious}
\end{figure}

Regardless of the approach used, when an investigator obtains access to the iHealth cloud portal, they can recover user metadata as well as device readings taken from the scale. Concerning user metadata, the user's name, birthday, height, weight, gender, and location are all recoverable. Device readings are recoverable in terms of textual and graphical information. The iHealth portal produces charts that plot the user's calories and weight information. Moreover, the portal also lists the scale's previous usage measurements. Metadata related to the scale readings recovered from this list includes the date and time, the user's weight, and body mass index. Figure \ref{fig:portalprevious} displays an exert of the previous entries as documented in the portal.

\subsection{Manipulations Analysis}

Table \ref{tab:Manipulations} -- Manipulations Results summarizes the scale readings that can be recovered from each of the evidence sources, in each day of the experiment. While the detail of evidence recovered from each evidence source is different, four common artifacts can be recovered from all four sources: the user's name or identifier, the weight measurement from each time the user used the scale, the date of each measurement, and the time of each measurement. Prior to any manipulation being introduced, scale readings were recovered from all four evidence sources, across all four days. 

\begin{table}[h]
	\begin{center}
		\begin{tabular}{| l |c | c | c| c|}
			\hline			
			
			\multicolumn{1}{|c|}{\textbf{}} & \multicolumn{4}{|c|}{\textbf{Day}}  \\ \hline	
			
			\multicolumn{1}{|c|}{\textbf{Result Set}} & \multicolumn{1}{|c|}{\textbf{1}} & \multicolumn{1}{|c|}{\textbf{2}} & \multicolumn{1}{|c|}{\textbf{3}} & \multicolumn{1}{|c|}{\textbf{4}}  \\ \hline	
			
			Result Set 1 -- Android & \checkmark & \checkmark & \checkmark & N/A \\ \hline
			Result Set 1 -- iOS & \checkmark & \checkmark & \checkmark & N/A \\ \hline
			Result Set 1 -- Scale  & \checkmark & \checkmark & \checkmark & N/A \\ \hline
			Result Set 1 -- Cloud  & \checkmark & \checkmark & \checkmark & N/A  \\ \hline
			\multicolumn{5}{|c|}{\textit{1st Manipulation -- Apps Cleared and Deleted}} \\ \hline
			Result Set 2 -- Android & \sffamily {X} & \sffamily {X} & \sffamily {X} & N/A \\ \hline
			Result Set 2 -- iOS & \sffamily {X} & \sffamily {X} & \sffamily {X} & N/A \\ \hline
			Result Set 2 -- Scale  & \checkmark & \checkmark & \checkmark &  N/A  \\ \hline
			Result Set 2 -- Cloud  & \checkmark & \checkmark & \checkmark & N/A \\ \hline
			\multicolumn{5}{|c|}{\textit{2nd Manipulation -- Scale User Deleted}} \\ \hline
			Result Set 3 -- Android & \checkmark & \checkmark & \checkmark & N/A \\ \hline
			Result Set 3 -- iOS & \checkmark & \checkmark & \checkmark & N/A \\ \hline
			Result Set 3 -- Scale  & \sffamily {X} & \sffamily {X}& \sffamily {X} & N/A \\ \hline
			Result Set 3 -- Cloud  & \checkmark & \checkmark & \checkmark & N/A \\ \hline
			\multicolumn{5}{|c|}{\textit{3rd Manipulation -- User Deleted and New Reading}} \\ \hline
			Result Set 4 -- Android & \checkmark & \checkmark & \checkmark & \checkmark \\ \hline
			Result Set 4 -- iOS & \checkmark & \checkmark & \checkmark & \checkmark \\ \hline
			Result Set 4 -- Scale  & \sffamily {X} & \sffamily {X} & \sffamily {X} & \checkmark \\ \hline
			Result Set 4 -- Cloud  & \checkmark & \checkmark & \checkmark & \checkmark \\ \hline
			\multicolumn{5}{|c|}{\textit{4th Manipulation -- New Reading Deleted}} \\ \hline
			Result Set 5 -- Android & \checkmark & \checkmark & \checkmark & \sffamily {X}  \\ \hline
			Result Set 5 -- iOS & \checkmark & \checkmark & \checkmark & \sffamily {X}  \\ \hline
			Result Set 5 -- Scale  & \sffamily {X} & \sffamily {X} & \sffamily {X} & \checkmark \\ \hline
			Result Set 5 -- Cloud  & \checkmark & \checkmark & \checkmark &  \sffamily {X} \\ \hline
			
		\end{tabular}
	\end{center}
	\centering {Key: \checkmark = Reading Recovered; X = Reading Not Recovered; N/A = Not Applicable in Result Set}

	\caption{Manipulations Results}
	\label{tab:Manipulations}
\end{table}

In terms of the first manipulation, when the smartphone applications' cache is cleared, and the applications are deleted, the results indicate that evidence related to scale readings cannot be recovered from the Android and iOS applications. These results support previous findings \cite{grispos2015recovering}, that the recovery of deleted artifacts from smartphone applications is a challenge for the forensics community. However, while the analysis of the applications did not result in artifacts related to the user or their use of the scale, the cloud portal and the scale itself can be used as evidence source substitutes in this scenario. Both of these evidence sources contain artifacts related to the user and their use of the scale, when the smartphone application's cache is cleared and the applications are deleted. 

With regard to the second manipulation, when the user is removed from the scale, both smartphone applications and the cloud portal still contain evidence related to the user and their activities. However, the analysis of the scale's binary dump revealed that the user's profile and their readings are overwritten with zeros and can not be recovered. This suggests that the scale does not communicate with the iHealth cloud to `scrub' other evidence sources in the ecosystem. Hence, a forensic investigator can use the artifacts from the applications, as well as the cloud portal, to reconstruct events that have taken place when the user's data is removed from the scale. 

Concerning the third manipulation, a new reading is undertaken after the user's profile has been deleted, the analysis indicate that the Day 4 reading can be recovered from all four evidence sources. An interesting observation is that when the scale's binary dump was examined, the correct user identifier was appended to the scale reading. This suggests that the scale is, potentially, communicating with the iHealth service to verify the individual. An alternative theory is that the scale `verifies'  the user's weight, by comparing the current reading with previous readings and then determines which individual is currently using the device. The reading is then appended with the respective user identifier.  However, further investigation is needed to confirm both theories. 

In terms of the fourth manipulation, when the Day 4 reading is deleted from the smartphone application, the results show that the smartphone applications and the cloud portal no longer contain metadata related to the reading. However, the reading can still be recovered from the scale's internal memory. 

\subsection{Analysis Summary}

The above research findings can be used to provide answers to the research questions presented in Section \ref{method}. Metadata about an individual and their use of a IoT device can be recovered from its respective IoT ecosystem. The findings suggest that metadata about the user and their use of the scale, can be recovered from various components (i.e. evidence sources) within an IoT ecosystem. With regard to the smartphone applications, metadata can be recovered from databases and XML files, while user and device usage metadata can be recovered from the scale's internal memory. Moreover, the individual's username and password can provide an investigator with further metadata from the iHealth cloud portal. 

The metadata retrieved from the IoT ecosystem describes the user's activities and device usage. Metadata recovered from the HS6 scale, the smartphone applications, and the cloud portal include weight measurements, along with timestamp information describing scale usage. Furthermore, body water measurements, body fat measurements, muscle mass values, visceral fat ratings, bone mass values, and body mass index values can also be recovered from the evidence sources in the evaluated ecosystem. However, the recovery of metadata from the evidence sources within the scale ecosystem is affected by the user's manipulation of the scale and the smartphone applications. For the first manipulation, where the cache is cleared and the smartphone applications are deleted, the results show that metadata concerning the user and their medical activities is no longer recoverable from the smartphones, but can still be recovered from the scale and the cloud portal. Hence, removing the user's metadata from the smartphone applications does not influence the metadata stored on the other evidence sources. For the second manipulation, when the user's profile is removed from the HS6 scale, the results show that the user's profile and readings are overwritten with zeros and can no longer be recovered from the scale's internal memory. However, user and scale metadata can still be retrieved from the smartphone applications. With regard to the third manipulation, when a new scale reading is taken after the user profile has been removed from the scale, the results show that the new reading can be recovered from the scale, the smartphone applications, and the cloud portal. The analysis of the scale memory dump shows that even if the user's profile has been removed from the scale, the measurement is correctly assigned to the correct user's ID number. The same is also true within the smartphone applications. For the fourth manipulation, when the new measurement is deleted using the smartphone applications, the results show that the scale's memory dump is the only evidence source that can be used to recover the deleted measurement. Hence, deleting a measurement using the smartphone applications does not remove the measurement from the scale's internal memory.

The results from the experiment support the hypothesis proposed in the introduction. IoT ecosystems provide forensically--relevant residual data. This statement is true for all the evidence sources evaluated in the iHealth scale ecosystem. The metadata generated by the evidence sources within the IoT ecosystem could be used by a digital forensic investigator as evidence, should the need arise.  

\section{Conclusions and Future Work} 
\label{conclusions}

The amalgamation of IoT ecosystems into both personal and hospital environments has attracted the attention of cybercriminals. As a result, there is a growing interest from law enforcement and private sector digital forensic investigators, as well as academia to identify user information and device operation metadata generated and collected by IoT ecosystems, which could be submitted as evidence in court. This work  investigates approaches for extracting potential digital evidence from IoT ecosystems and evaluated the residual data present in the ecosystem when different manipulations are undertaken involving the various components of a scale IoT ecosystem. The potential evidence recovered from these components could provide investigators with information such as user's medical history and well--being, at a particular time. This data could be critical when investigators need to create a timeline of events that highlight when a particular victim has died, and what factors could have contributed to their death. 

There are several potential research opportunities. For example, one future research agenda is to investigate a greater variety of IoT ecosystems (e.g., smart security, smart home or Industrial IoT (IIoT)) and the components that make up these ecosystems. The investigative approach presented in this paper can also be extended to other IoT devices from a variety of manufacturers. In addition, the smartphone devices included as part of the scale IoT ecosystem can also be extended to include smartphone applications that run on other mobile operating systems. Another separate avenue of research can examine the extent to which techniques in this paper can be extended to traditional networked devices, such as infusion pumps, user monitoring devices, smart meters, and industrial control systems. The purpose of this research is to evaluate the ability to recover user and device metadata from the internal memory of other devices. The results from this research can also be used to investigate if device development should integrate forensic--by--design principles, in order to enhance forensic investigations for IoT devices.

\bibliographystyle{IEEEtran}
\bibliography{IEEEabrv,cas-refs}
\enlargethispage{-5cm}
\end{document}